# AN OLD-FASHIONED FRAMEWORK FOR MACHINE LEARNING IN TURBULENCE MODELING

Philippe Spalart

Woodinville, USA

prspalart@gmail.com

**Abstract**

The objective is to provide clear and well-motivated guidance to Machine Learning (ML) teams, founded on our experience in empirical turbulence modeling. Guidance is also needed for modeling outside ML. ML is not yet successful in turbulence modeling, and many papers have produced unusable proposals either due to errors in math or physics, or to severe overfitting. We believe that "Turbulence Culture" (TC) takes years to learn and is difficult to convey especially considering the modern lack of time for careful study; important facts which are self-evident after a career in turbulence research and modeling and extensive reading are easy to miss. In addition, many of them are not absolute facts, a consequence of the gaps in our understanding of turbulence and the weak connection of models to first principles. Some of the mathematical facts are rigorous, but the physical aspects often are not. Turbulence models are surprisingly arbitrary. Disagreement between experts confuses the new entrants. In addition, several key properties of the models are ascertained through non-trivial analytical properties of the differential equations, which puts them out of reach of purely data-driven ML-type approaches. The best example is the crucial behavior of the model at the edge of the turbulent region (ETR). The knowledge we wish to put out here may be divided into "Mission" and "Requirements," each combining physics and mathematics. Clear lists of "Hard" and "Soft" constraints are presented. A concrete example of how DNS data could be used, possibly allied with ML, is first carried through and illustrates the large number of decisions needed. Our focus is on creating effective products which will empower CFD, rather than on publications.

# Introduction

Turbulence modeling has a mixed reputation. Many observers regret how far it is from perfection and describe a sense of stagnation, which by one simple measure goes back to 1992 [1, 2]. The community's expectations have risen considerably since then. Very few people are active in the creation or improvement of models, compared with those engaged in CFD code improvement, some of them for profit; the investment into computers is also massive. Some major early contributors to modeling have withdrawn to neighboring fields.

This is concurrent to the relentless progress in numerical accuracy. For decades, it was propelled by Moore's Law; nowadays it rests just as much on progress in strong high-order solvers and grid adaptation. Whereas for many years, turbulence-modeling flaws could not be conclusively distinguished in practical work or even in semi-complex workshop cases from lack of numerical convergence, we are nearing the era of negligible numerical errors for steady-state flow solutions, and this even for complex geometries such as an airplane wing with multiple slats, flaps, and their supports [3]. Secondary sources of error such as elastic wing and flap support deformation and wind-tunnel effects are also better controlled. This will leave as the only culprit the errors associated with turbulence treatments. They will be finite, and larger than what would be acceptable in industry.

Anticipating this situation suggests increased strategic funding for modeling and a look for alternate approaches to that in force since the 1970s. Machine Learning being so successful in other fields, it is an obvious candidate, and the last 5 to 10 years duly have seen a massive rise in activity and publication [4-6]. However, some paper titles exhibit hubris and a grave ignorance of how hardened the turbulence problem has become [7]. This activity certainly has not had immediate success. To the author's knowledge, none of it has produced a new model or model version that has been inserted into the Turbulence Modeling Resource [8], NASA and similar solvers, or commercial CFD products. There were no signs of that at the 2022 High-Lift Prediction Workshop [9], and the NASA workshop confirmed this situation [10]. Complacency is possible in government and commercial actors, but this is revealing. Very few of the ML papers produced models which could be coded by others, or which had clear hopes of being general. Most often, the study is "a successful ML project" but fails to provide an appropriate "product." This is especially true when the model violates a very well-established rule of Reynolds-Averaged Navier-Stokes (RANS) modeling such as Galilean Invariance. Such incidents were a motivation for the present paper.

International funding and collaboration have been quite good for the production of turbulence data, from experiments and simulations. The community is exhibiting intelligence and goodwill. Funding for the classical modeling activity itself has been much smaller, but the limiting factor here is more our lack of talent than a lack of will in the agencies. Unfortunately, so far, the sustained production of detailed data from experiments and DNS of the last decade has had essentially no direct effect on models; an exception is the Quadratic Constitutive Relation calibrated from Juncture-Flow experimental data [11]. No matter how good it is, a new dataset only adds a little to the large set of flows the model is attempting to address. Collectively, we have failed to achieve this cross-fertilization.

*Turbulence-Resolving Simulations*

A concurrent debate takes place over the emergence of Turbulence-Resolving (TR) simulations, and the claim in some quarters that they are now supplanting RANS modeling in terms of accuracy and even cost. A common and rather valid complaint is that "traditional turbulent CFD cannot predict separation." It helps to divide this complaint into the poor prediction of incipient separation from a smooth surface, and poor predictions for massive separation. For the latter, we are close to a consensus that steady RANS indeed has very little chance; take the example of a circular cylinder. A model of any complexity which in a steady setting accurately reproduces the flow, averaged over cycles of Karman vortex shedding with their three-dimensionality, large-scale time-dependence, and modulations, would be amazing. In such situations the argument in favor of TR is very strong. On the other hand, for smooth-body separation, the only practical LES approach from a computing-cost point of view at least in Aerospace is Wall-Modeled Large-Eddy Simulation (WMLES), which remains complex, ever-changing, and controversial. Recent comparisons between WMLES and DNS on the Speed Bump, which is a demanding case of smooth-body separation [12], have not been favorable.

A key obstacle for WMLES is the very large number of degrees of freedom (DOF) required to resolve the thinnest region of the boundary layer (BL). It hinges on the number of cubes needed to fill the boundary layer, denoted by $N_{cubes}$ [13]; it is the integral over the surface of $1/\delta^2$, where $\delta$ is the boundary-layer thickness. This concept was introduced in 1997, but sadly the WMLES proponents studiously avoid reporting it, or discussing its meaning to their line of work.

The excessive number of DOF needed for an accurate WMLES over a full-size wing, for instance, has led since 1997 and the introduction of Detached-Eddy Simulation (DES) to wide-spread work on Hybrid RANS-LES Methods (HRLM) [14]. In these, RANS is used in the boundary layer, often all the way to the separation line (or trailing edge, or other sharp edge), and LES thereafter. HRLMs have decades of life ahead of them, and therefore, vigorous efforts in RANS modeling, particularly of boundary layers, remain highly justified. These hybrid methods are also complex and imperfect, but versions of DES are widespread and give close answers in different solvers, so that a knowledge base shared by many actors now exists.

Note that ML has also been applied to WMLES, and the boundary between the WM component in WMLES and RANS is porous, although WM is both more involved, since the interaction between the very-near-wall region and the outer part of the boundary layer is very active and impactful, and more limited in range. Wall Models in that sense will not be covered here.

Some practical considerations are unfavorable to TR approaches. The simulation starts from a poor initial condition, runs for a settling time $T_1$, and is averaged over an interval $[T_1,T_2]$. These times are difficult to choose (although Mockett et al. made an attractive proposal [15]). They are necessarily questionable, and the error decreases only proportionally to $1/\sqrt{T_2 - T_1}$, which is very slowly. This creates noise from averaging, which appears to completely rule out optimization, inverse design, or adjoint operators. These are major components of modern CFD. In other words, TR work is for now restricted to flow analysis.

*Other remarks*

A remark which will appear self-serving is that, stagnation aside, it is surprising that RANS models with so few equations serve the engineering practice quite well. One reason is that this good service takes place overwhelmingly in boundary layers, which are dominated by a single quantity, a Reynolds shear stress (with additions for 3D boundary layers and heat transfer). The trend towards HRLMs may support the idea of creating models which emphasize boundary layers over any other flow type, provided they do not cause severe damage in such flows. This has not been tried yet. In any case, offering models with few equations and a moderate impact on the solver's convergence (both versus grid resolution and iterations) very much meets the desires of the engineering community.

The resilience of models with at most two equations implies a relative disappointment for Reynolds-Stress models and their seven equations in terms of accuracy. In recent years, increased computer power, symbolic manipulation capabilities and detailed DNS data could have led to the adoption of models with even more equations, which should have given higher accuracy as a natural consequence of systematic model development. The Closure Problem, which essentially prevents this, is a non-trivial part of Turbulence Culture. Some groups also report chronic difficulties with numerical convergence for the higher models.

This paper could have been written in 1990, when the author was constructing the Spalart-Allmaras (SA) model [1]. He did not expect the long career of the model, which gives him some authority to write now. It is disappointing that one of the key requirements discussed here, namely the ETR, was presented 30 years ago by SA and expanded to two-equation models by Cazalbou, Spalart & Bradshaw [16] in 1994 (a Physics-of-Fluids paper cited only 91 times), but even now is not understood by most of the creators of new models. It was called the Turbulent-Non-Turbulent Interface (TNT) by Kok [17]. A refinement of the external field for SST was proposed, but is not used much [18]. The two dominant models in Aerospace are due to authors, namely SA and Menter for SST, who clearly identified and explained the importance of the ETR. This is why below we will insist on it, again.

A minor remark is in order, regarding the $k$-$\epsilon$ model. It was established in the 1970s, and ETR considerations probably played a role in the choice of the second variable. Its prestige remains high, long after its low accuracy for adverse pressure gradients and separation has been demonstrated, and repeated efforts to resolve that have failed. The SA and especially the $k$-$\omega$ models are preferable in boundary layers; we do not know of any explanation of that trend from first principles. Yet many researchers and users consider $k$-$\epsilon$ as the first one to try, especially in upgraded versions such as RNG, Realizable, or nonlinear. This could represent a negative form of Turbulence Culture.

Another example of negative culture is the concept of "equilibrium" and "non-equilibrium" turbulent flows, which is simplistic and much over-used [19]. Related to it is the endless focus on the fact that the production and dissipation would be equal. This is meaningless for the most important Reynolds stress in thin shear flows, namely $\overline{u'v'}$, because the dominant term opposing production is the pressure term rather than the dissipation [19,20]. The "self-evident status" of the condition "$P = \epsilon$" must be destroyed.

A side note is that more than once, the author attempted to develop a two-equation model with the same nature as the SA model, that is, directly by reflection over the eddy viscosity instead of seeking an approximation of exact equations such as the one for Turbulent Kinetic Energy (TKE). The SA model "has no ancestors" [1]. Each time, the author was unable to think through the countless options he faced, for instance regarding the variables: $k$, $\varepsilon$, $\omega$, $\nu_t$, combinations of those, or a new idea? A similar challenge developed when he attempted to devise a very general Quadratic Constitutive Relation, beyond the simple first proposal which has only one term [21]: it rapidly became clear that ten or more constants would be needed, which stopped the attempt. The QCR2020 model is much simpler [11]. In other words, the idea has merit that the brain-power challenge of creating new models or even new corrections is major.

*The Precise Role of ML*

The role of AI and ML in turbulence modeling could take different forms. At one extreme, ML replaces the "natural intelligence" of the traditional modelers thanks to its superior power relative to a human brain. It becomes the architect of the model. The principal task at that stage is to choose which quantities enter the model, among countless candidates such as derivatives of the flow field and model quantities. This "design space" is unbounded. We know of no clear example of this being achieved through AI.

At the other extreme, ML is only a tool for subtasks conceived by humans. The simplest would be to optimize a constant, say $c_\mu$, over a group of cases. Another example would be the $f_{v1}$ function in SA [1]. The 1992 model uses a simple formula due to Mellor & Herring. Nowadays, Direct Numerical Simulations (DNS) provide a more accurate $f_{v1}$ distribution, and a properly constrained Neural Network or search for analytical expressions would give a "better SA model" for flows with weak pressure gradients. The ML work would, fashionably, be "data-driven" and lead to "near-perfection." This has not been done and the primary reason was to avoid creating a new version of SA, which is costly to the community, for only a very small benefit. In any case, in this extreme the human has still fully determined the argument of $f_{v1}$, namely $\chi$, and its role in the model. A preliminary attempt to improve the similar $f_w$ function instead is described below.

*Outline of the Paper*

The body of the paper is divided into considerations of Mission and of Requirements, following a technical exploration of the "Mechanics" of ML. The Mission discussion is the less specific one and somewhat "in the eye of the beholder," the way Beauty is, but deserves careful attention because a clear vision of how models contribute to the engineering community is lacking in many papers. The core of the Requirements section is a "check list" of specific characteristics a model must or should have. Many of these are well-motivated, by physical reasoning and also by the "marketplace" for models, but are not unquestionable. Therefore, a discussion will be provided for each of them.

## The Mechanics of DNS Use in Turbulence Modeling

We will call Training in Real Space (TRS) the ML approach that has been almost universal. Accurate data are obtained for a flow, in a region in *(x,y,z,t)* space. Averaging typically reduces the dimension from 4 to 2 or 1. Then, the principle is to improve the model's predictions, for instance of the mean flow or Reynolds stresses, over this region. A common tool is Neural Networks (NN), but Gene-Expression Programming is also used [22], as is manual fitting. An existing model can be improved, or a new model can be calibrated. An obvious danger with a single-flow study is of overfitting over a very small part of the "Turbulence Universe."

We present in detail an example for the SA model, which appears simple at first but encounters the key concepts. We take channel flow, for which high-quality DNS databases are available. In such a flow with a single curve, ML is not truly needed, and a manual fit would be practical. The situation is different in other flows. The exercise is still surprisingly complex.

The domain is one-dimensional, in *y*. Outside the viscous region the SA transport equation reduces to:

$$c_{b1} U_y \tilde{v} - c_{w1} f_w \left(\frac{\tilde{v}}{y}\right)^2 + \frac{1}{\sigma}\left[\partial_y(\tilde{v}\tilde{v}_y) + c_{b2}\tilde{v}_y^2\right] = 0. \tag{1}$$

The DNS produces the eddy viscosity $v_t$ as the ratio of Reynolds shear stress to shear rate $v_t \equiv -\overline{u'v'}/U_y$, rather than the SA variable $\tilde{v}$, but the near-wall viscous and buffer region in which they differ significantly is excluded, and the notation $\tilde{v}$ will be used. Once $\tilde{v}_{DNS}$ is extracted and inserted into (1) all the terms are known, and of course the equation is not exactly satisfied by the SA model (including the original $f_w(r)$ function), applied to the DNS results.

We then consider the $f_w$ function as the candidate for improvement (rather than, for instance, the constant $c_{b2}$, or inserting a correction to the production term as other teams have decided to). We calculate the "DNS distribution" from (1) and simple algebra, and call it $f_{wDNS}$. It and $r$ are shown in Fig. 1a; recall that $r \equiv \tilde{v}/(U_y \kappa_{SA}^2 y^2)$ and a key assumption in the SA model is that making $f_w$ a function of $r$ only is a reasonable approximation. Many findings ensue.

We must mention that we did not use all the DNS datasets that are available. Some of them give for $f_w$ results that are so different from the three here, which come from three different groups, that it would be useless to further obscure the figures. In a definitive paper on this topic, the discrepancies would need to be explored.

In Fig. 1a both quantities start near 1. This is the first finding: it indicates that the $\kappa$ value set for SA, namely 0.41, is at least in fair agreement with the DNS (observe that $r = (\kappa_{DNS}/\kappa_{SA})^2$). $f_w$ and $r$ then come down: they are correlated, as was assumed when creating SA. In agreement with the classical scaling, here the Defect Law, the quantities are independent of the flow Reynolds number $Re_\tau$ when normalized by $u_\tau$ and $y/h$. An additional finding is that $r$ is monotonic at first but not near the centerline where it tends to $\infty$ because $\tilde{v}_{DNS}$ is finite and $U_y$ falls to 0; the behavior in the boundary layer is better.

This leads to Fig. 1b, in which $f_{wDNS}$ and $f_{wSA92}$ are shown versus $r$, displaying a substantial disagreement. Channel flow was not used in the calibration of SA. Then comes the usual challenge of progressing from a flow-specific distribution $f_{wDNS}(y/h)$ to a model $f_w(r)$. Potentially, one would fit a simple NN to $f_{wDNS}(r)$, but only after excluding the buffer and the "0/0" region near the centerline, which here means any $r < 0.4$. Thus, meaningful decisions are needed.

Several pitfalls of TRS are in evidence. As mentioned, the first would be to propose the function $f_w(y/h)$ in Fig. 1a as a "new model." Quite a few ML papers have had as their output such corrections, often denoted by $\beta(x, y)$ for instance. This ignores the "universality principle" and the requirement for turbulence models to use local quantities unrelated to the geometry. Recall that both the SA and SST model use $y$ in their equations, based on empirical arguments of wall-interference, but not $h$.

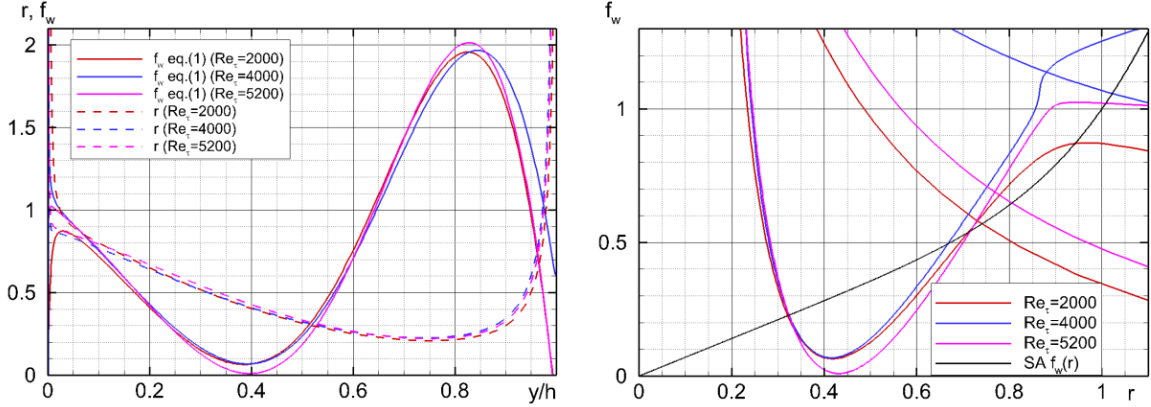

Fig. 1. The $f_w$ function in channel flow, trained by DNS. DNS data by Hoyas, Pirozzoli, and Lee & Moser. Courtesy A. Garbaruk.

The second pitfall is overfitting, since $r$ only covers [0.22,1] and its useful range is closer to [0.4,0.85] whereas $f_w$ must be available over $[0,\infty[$. The behavior of the NN outside its training interval is unpredictable. The disagreement for $r > 0.85$ is curious, considering how well DNS from different sources agrees very near the wall. It probably is amplified by the various derivatives and divisions present in (1).

The third pitfall is that arrangements for troublesome "0/0" phenomena require human intervention; thus, the work is not "pure ML." This problem is even worse in external flows, which have inviscid regions in which many turbulence quantities are "0/0;" those regions should not influence the model. The approach of weighting different regions intentionally is again possible but depends on a deliberate intervention inside the penalty function. A regularization is justified in the Cost Function, but is also arbitrary. The simplest strategy would be to create a cost uniformly distributed over [0,1] for $y/h$, but a better one would weight the wall region more heavily. Recall that in the SA paper, the slope was adjusted to obtain the right skin friction in a boundary layer, and the channel was not considered. Note also that the 92 model has $f_w(0) = 0$, but there is no absolute reason for that; the best behavior could be a finite value, or conversely be $O(r^2)$.

Other comments are in order. To extract $f_w$, Dr. Garbaruk had to smooth the eddy-viscosity distributions by fitting fifth-order polynomials to them; those coming from the raw DNS data were too noisy, especially since (1) involves second derivatives. This will be a common problem in TRS from DNS, not to mention experiments.

In channel flow, the eddy viscosity has only one possible definition. In 2D and 3D flows, a definition of the "effective eddy viscosity" in the DNS by a least-squares fit has merit and is widely used, but it is an empirical concept. Of course, in such flows, a scatter plot of $f_w$ versus $r$ will be a cloud rather than a single curve, and the first question will be "are the two quantities strongly correlated?" For the velocity log layer to fit, we need the condition $f_w(1) = 1$, so that an ML fit would have to be constrained; this is another deviation from a "pure ML" operation.

The overall conclusion from the SA/Channel DNS exercise is that the common expectation that ML will work "by itself" is erroneous. Model improvements involve a large number of decisions (in addition of course to the hyper-parameters of the NN) and demand an intimate knowledge of the model. The other conclusion is that our strategy for improving $f_w$ has not concluded so far. At the very least, a flat-plate boundary layer will be needed.

Models with more than one equation complicate the situation. It is essential to obtain the correct eddy viscosity, which in the *k-ω* model is $\nu_t \equiv k/\omega$. However, it is not the case that $\nu_{tDNS} \equiv k_{DNS}/\omega_{DNS}$, even outside the viscous region. The way the models work is that is that $\nu_{tRANS} \equiv k_{RANS}/\omega_{RANS}$ takes correct values, but its separate components do not. For instance, $k_{RANS}$ is constant in the log layer, and therefore cannot reproduce $k_{DNS}$; the latter has a slope in the log layer, and a dependence on $Re_\tau$. This is part of the "Structural Limitation" (SL) for such models, and also for Reynolds-Stress Models (RSM), as explained by the author in a separate paper which presents the strong conjecture that there exists a Generalized Law of the Wall (GLW) for Classical models [23]. The GLW and SL predate ML, of course. There has been no proposal to address this deep issue, even as considerable effort is spent generating DNS databases for training purposes [24]. The only immediate option appears to be to train the models based only on mean-flow predictions, instead of the entire Reynolds-stress tensor (or even its trace *2k*) and dissipation rate ε. However, this goes against the valid desire to involve the physics of turbulence. The author's belief is that it is urgent for the Structural Limitation to be recognized, while he has been unable to suggest any solution to it within the established structure of models.

# Mission

The mission is to provide well-understood, stable, and decently accurate approximations to the Reynolds stresses and heat flux inside a Navier-Stokes solver, at low cost.

There are several aspects to the true cost of treating turbulence, besides the number of equations, just like a Navier-Stokes solution is more costly than a potential-flow solution. One is the number of grid points in the boundary layers; another is the damage to the iterative convergence rate. In steady turbulence solutions, the number of points in the inviscid and the turbulent regions is of the same order. Regarding the number of equations, going from one to eight PDEs in addition to the five of the q-vector is manageable, but proposing a Neural Network (NN) with ten layers of ten neurons each may not be.

We specify the Reynolds stresses and heat flux as the products of the model, although only combinations of their derivatives enter the mean-flow equations, because proceeding through the stresses and flux is essential to conserving momentum and energy. A beginner in modeling could attempt to directly model the derivatives, as the author saw in a publication a week ago, and the fact that this would be unacceptable may be a first example of a "self-evident" fact in TC.

A valid question of Mission is whether all six Reynolds stresses need to be accurate. As mentioned, it is common for models to produce an inaccurate Turbulent Kinetic Energy $k$ but a correct eddy viscosity and therefore a correct Reynolds shear stress, which is the only one influencing the mean flow in boundary layers (at low Mach number). Still, Reynolds-Stress Models and even Non-Linear Eddy-Viscosity Models especially in a "next generation" developing into the 2020's, informed by DNS and ML, will widely be expected to improve all the stresses. If the GLW-based Structural Limitation conjecture is correct, a complete success would be difficult [23].

## *Universality*

An essential aspect of Mission is Universality, or Generality. It is not a rigorous demand, but it has a long history. The argument is that turbulence physics is very similar in most of the flows of engineering interest, and imperfect as it is, a single model is offered for all flows. The model would generalize thanks to built-in invariance and physically correct inputs, in other words a sound design (and maybe a little luck). Now the mainstream models, after decades of service, have been given

"corrections" in addition to the core formulation, for example to address compressibility or surface curvature. But the core model and all its constants have been static. It is good practice for the custodian of a model not to propose marginal modifications, because the community will then run both versions, thus cluttering the knowledge base in workshops and elsewhere, and adding cost to the already considerable collaborative-testing effort.

A clear statement from the pioneer years of differential modeling, which we strongly agree with, is by duP Donaldson (following work with Rosenbaum) [25]: "It is our thought that the simpler the model is the more general will be its applicability, and it is generality that is our primary goal." A substantial advantage of simplicity is that it eases the understanding and improvement of the model.

A valid argument for Universality is that modern CFD is capable of solutions for flows containing a wide variety of "flow modules," and a single model is asked to provide predictions of "decent accuracy" in all these modules. An airliner in landing configuration presents with, at least: simple but long boundary layers; BL in 3D flow; BL with curvature; BL in pressure gradient up to shock waves; complex 3D obstacles causing flow reversal, both small and large like the landing gear; wakes possibly merging with the BL; shear layers from the trailing edge or the jet engine; separation from sharp edges; vortices free or embedded in the BL; cavities; horseshoe vortices; corner flows with weak turbulence-generated vortices. The idea of a zonal model explicitly controlled by the user to optimize it in each area is very unattractive in industry.

### The GEKO Model

A recent major challenge to this universality principle is the GEKO model established by Menter in the ANSYS CFD solver [26]. The model is adjustable in six directions, the most important one being the prediction of separation. Based on their experience, the users can instantly dial a model with early or late separation, on either side of the SST model which is viewed as the best compromise in the absence of evidence against it in any particular class of flow. A wind-turbine engineer will run a different model from the auto engineer. The settings can be different in different flow regions: the model is zonal in just the way we described as "unattractive." This controversy will play out in the near future.

Another deviation from common turbulence-modeling practice is that the GEKO equations are not made public, at least for now. This is a business decision; the model will not be coded in other solvers, which greatly reduces the mass of the knowledge base for it. This is a step in the opposite direction of the Open-Source model, which has spread through the AI community for other lines of work. Until now, CFD solvers have been kept proprietary even if the algorithm (for instance multi-grid) was published, but turbulence models have been in the open.

An essential feature in the conception of the GEKO model is that the predictions of the flat-plate boundary layer are unchanged through the six adjustments. The present author has applied exactly this principle in the few unpublished versions of the SA model. It reflects the power of this flow module to set the boundary-layer thickness, which then influences separation once a pressure gradient intervenes. The skin-friction drag is also of great importance especially in Aeronautics. This is a comment on how the simple boundary layer is the one undisputed area of success for turbulence modeling.

In ML and other work, the important consequence of this "Universality Principle" is that publishing a very narrow model, for instance one for jets in a cross-flow, is of questionable value, especially if it destroys the accuracy in boundary layers.

*Possible Machine Learning Strategies*

In theory, a framework to develop universal models by ML would be straightforward. A Cost Function (CF) would be formulated that incorporates the error over flows in all the classes of interest, and then the CF be minimized which would produce the best compromise. This evokes the use of "Big Data," although the size there has more often referred to that of the physical domain over which the error is calculated than the number of flows, often measured by the number of grid points. The intellectual effort of choosing flows and for each one setting a region of interest and an overall weight in the CF would be considerable. In comparison, the SA92 model rested on "Turbulence Facts" countable on the fingers of one hand. Additional decisions would be between an exercise that only adjusts constants such as $c_{b1}$, or provides functions possibly through Neural Networks, or chooses analytical forms possibly by Gene Expression Programming [22]. In any case, the computing effort would be out of reach if, as most people expect, two-dimensional fields were included. Calculating gradients to conduct the optimization also appears very difficult.

## *Publishing Practices*

We turn from the mission of the model to that of the modeler. We assume the traditional model for publication, motivated by a passion for knowledge, community service, and prestige. Over the life of a successful model, the initial cost of creation is negligible; the financial cost of each CFD run is borne by the users themselves; the cost of the knowledge base that outlines the qualities and the limitations of the model is taken on by the community, the workshop organizers, the publishing industry, and the generous reviewers. The success of a model, which is set after a few short years, is quite unpredictable; the key is for it to be tried by experts outside the team, and to do relatively well. Creating a truly new model involves technical risk. The reputation of the authors and their institution matters. Timing can help, if the evolution of CFD opens an opportunity for larger calculations or makes the need more pressing for a new capability, for instance compatibility with unstructured grids [27].

Some qualities are essential in a new turbulence model paper. The model equations must be absolutely complete, with clear notation. Traditionally, the equations are written on the page, but in the future a NN could be part of the model and be provided in one way or another. A number of specific Requirements will be presented in the next section, for instance the needs at the ETR. The input quantities, such as derivatives, should be plausible (we admit that this is a vague word). Good models respect basic aspects of TC such as dominance of large eddies away from walls, leading to a very weak effect of the molecular viscosity. The possible role of the wall distance and wall-normal vector must be clear. It is essential to provide the boundary conditions for the Partial Differential Equations. These include conditions at solid walls, at planes of symmetry, and at the inflow boundaries for external aerodynamics. Often, these are different when the intent is for any boundary layer to rapidly become turbulent, or it is not, and trips will be provided. Initial conditions which can improve convergence are desirable. Numerical solutions with widely different resolutions giving the same answer will greatly increase the confidence of potential "customers." The inability to achieve grid convergence is not unusual, and it dooms a model.

Other aspects of this initial paper are more debatable. It must be known whether the model is specialized or is general, with hopes for a career such as that of $k$-$\epsilon$. It is desirable to give a sense of the physical and mathematical reasoning behind the various terms, so as to reduce the impression of a "black art," and to reveal the debt to previous work. The most difficult part is to decide how many different flows to present results for. Basic cases such as the boundary layer are obvious, but more delicate flows such as shock-induced separation raise the level of effort. Nowadays, relatively complex 3D cases could be considered unlike in 1992. Stable results in these would be signs of robustness. There is a conflict between "going public" relatively rapidly with creative work, possibly receiving useful advice, and gathering the "critical mass" to have the attention of readers outside your circle, who are overwhelmed with an excess of information both at conferences and in archival literature.

Over the years, the responsibilities of the model's "parents" or custodians include version control and compatibility between "updates" and between the independent corrections, say for compressibility and for curvature [8]. Advice as to which corrections are desirable in a particular flow should be clear. If an update definitely supersedes an earlier version in the modeler's mind, this statement must be put on record. A proliferation of versions, which diligent users and workshop participants will compare, can cause a large waste of effort. The urge to publish should be resisted.

Few of the considerations here are related to ML. There is the danger of viewing a large dataset on a few flows to be "big data" and rushing to propose a narrow-based model. There is also the question of "how a NN is published." Another issue is that many ML procedures by-pass physical reasoning and intuition, so that the model is impossible to "understand" and very difficult to improve, unlike the traditional ones. Unfortunately, in the author's opinion, a very substantial issue for the "ML Generation" has been the use of quantities, such as the velocity magnitude, which are unacceptable. The papers introduce these quantities casually, well into the paper, as if this was not the extremely important characteristic of a model it actually is.

## Requirements

This section will be more concrete than the last one. Its title could be "Constraints." There is a need to clearly identify errors in the formulation of a model which make it hopeless as a general model; unfortunately, many of these errors are found in the ML literature, and will be made over and over. The non-ML literature is not immune. Flawed papers are accepted due to the shortage of competent reviewers with sufficient TC, and the modern unwillingness to seek rigorous mathematical facts at the expense of hours of algebra. Once a fallacious "theory" has been printed by a prestigious journal, it has status and propagates; we essentially face the danger of an "Alternate Reality in Turbulence" establishing itself. Pre-ML TC contains quite a few fallacies, old and new, supported by many publications [19].

As we did in earlier oral presentations, we distinguish "Hard" and "Soft" constraints; for both, there is a progression from obvious simple ones to some that require more algebra and reasoning, but have a considerable impact, an example being the ETR. For some of them, the author has been unable to convince some of his correspondents in private communications. His 2015 paper "Philosophies and Fallacies in Turbulence Modeling" explicitly disputes many of the common lines of thought but is cited fewer than 20 times a year; its impact is disappointing [19].

*Hard Constraints*

We first make a list of the simplest constraints, which should have been obvious. Again, we are aiming at a general model, which could be shared, for instance between work on a golf ball and work on a wind-turbine farm. For more precise reasons, we refer to the Fallacies section of the previous paper [19].

1. Respect dimensional analysis and tensor symmetries.
2. The equations must be well-posed. Do not propose a PDE by which short waves grow rapidly in time.
3. Do not use the mean velocity in the model itself.
4. Do not use quantities derived from the mean velocity itself, such as streamline curvature, helicity, or a gradient "in the flow direction."
5. Do not use properties that depend on the direction of the coordinate axes.
6. Do not use properties which depend on the averaged flow being steady.
7. Do not use the acceleration or the pressure gradient.
8. Do not expect the user to manually choose different models within a Zonal Method.
9. Do not use properties of the current flow problem, for instance the freestream velocity or the Reynolds number.
10. Do not propose a model without strong evidence that it behaves well at the Edge of the Turbulence Region.

Item 1 is unquestionable, and only pure errors have made some publications fail the test. A minor violation of Dimensional Analysis is common, when a model uses, for example, $\max(k, 10^{-14})$. What is really written is $\max(k, 10^{-14} m^2/s^2)$, or other units. Thus, a dimensional parameter has been introduced. This will be a problem only in exceptional situations.

Item 2 is a problem that is not always easy to detect by reading the equations. The simplest type is that of a negative diffusion coefficient. However, running a strong grid-refinement exercise will reveal it; by this, we mean refinement by about a factor of 2 in each direction, performed twice. Unfortunately, quite a few modeling papers have failed to do so.

For Items 3 and 4, observe that the velocity is part of the convection term, of course, but only combined with the time derivative $\partial/\partial t$ in the **same** reference frame to form the Lagrangian derivative $D/Dt$.

Item 5. In particular, the distinction between "normal" and "shear" Reynolds stresses is meaningless. Statements such as "the normal stresses are equal" also are meaningless, but quite common. The confusion comes from the frequent study of simple shear flows with *x* the flow direction and *y* the direction of variation.

Item 6: physically, only Lagrangian derivatives $D/Dt$ have any meaning. Put another way, a body of turbulence that crosses a deformation such as a shock wave is itself not steady in any sense, whether or not there exists a reference frame in which $\partial/\partial t = 0$.

Item 7 has been particularly difficult to impress on colleagues, due to the acceleration being Galilean Invariant and the pressure gradient being so dominant in discussions of boundary-layer transition and turbulence. But the situation is clear only with stationary walls, and a model should apply outside such cases.

A concise statement of the argument is as follows: knowing a solution $U_0(x, y, z, t)$ to the incompressible Navier-Stokes equations (capitals denoting velocity vectors), the flow field $U_1 = U_0 + \Delta U(t)$ is also a solution (with the independent spatial coordinates properly transformed); it will have the same turbulence. However, it has a different acceleration. Therefore, a model which uses acceleration and is accurate if written in terms of $U_0$ will be inaccurate written in terms of $U_1$. This was explained in 1999 by the author and Prof. Speziale [28], and by others, but that paper has been cited only 25 times.

Item 8 is a strong point, except in highly simplified geometries. Modern CFD is relentlessly progressing towards very complex geometries. Think of an automobile simulation with all the details under the hood and in the cabin. However, the GEKO model goes against it as explained above.

Item 9 is linked to the generality principle. Besides, the Reynolds number of a wing can be based on its span or its chord, on the freestream velocity or the speed of sound, and so on. These are not relevant locally. If $Re$ is a valid Reynolds number, so is $2Re$. A worse although rare problem is when the inflow boundary condition for the turbulence variables involves the speed of sound; any model must be applicable to incompressible flows.

Item 10 is crucial and is often ignored. Cazalbou et al. (CSB) showed that fairly simple analytical facts can be determined [16]. For example, Fig. 2 shows the complete set of quantities in the $k$-$\epsilon$ model. The ramp is propagating to the right, relative to the fluid. The eddy viscosity falls to 0 linearly, and the other quantities follow higher, non-integer powers of the distance to the interface. CSB also provided simple inequalities between quantities such as $\sigma_k$ and $\sigma_\epsilon$ that need to be satisfied for stable behavior and grid convergence when the values outside the turbulent layer are very small or 0, and of course for the ramp to propagate into the non-turbulent fluid. They predict the flaw in the $k$-$\omega$ model which was explained by Menter, and they were used by Kok to propose a healthy version of it, which would have been simpler than the SST model [17].

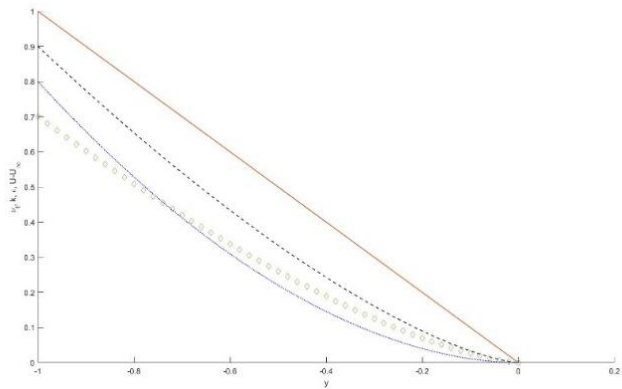

Fig. 2. Ramp solution for $k$-$\epsilon$ model. Solid, eddy viscosity; - -, k; ⋯, $\epsilon$; ◊, velocity. Arbitrary scaling for each quantity.

If even the CSB mathematics are too difficult, running the flat plate with extreme grid refinement, as is done at the TMR [8], will detect several of the possible flaws in a model. Running a simple airfoil will reveal possible unphysical behavior in its wake. Typically, it is that the turbulence falls to 0 in a finite distance, in conflict with classical theory and dimensional analysis, which indicate that the peak eddy viscosity goes to a finite constant.

The analysis is relevant to other models. In some models such as Baldwin-Barth [27], the ramp similar to Fig. 2 propagates to the left because $c_{b2}=-2$ (using SA notation [8]), and this is a major problem; we believe it makes grid convergence impossible, and the problem is hidden by not performing enough grid refinement. This is illustrated vividly by analytical solutions for the SA and the Baldwin-Barth models, shown in Fig. 3. These are solutions of only the diffusion terms, thus representing the evolution of a layer of turbulence, stirred at the initial time, and left to decay and propagate in the absence of shear and therefore production (the solutions also work out in cylindrical and spherical coordinates). The solutions are parabolas:

$$\nu_t = \nu_{tmax}(t)\left(1 - \frac{y^2}{\delta^2(t)}\right),$$

for $|y| < \delta$ and 0 outside, and the diffusion equation leads to:

$$\frac{\nu_{tmax}}{\nu_0} = \left[1 + \frac{2\nu_0(3+2c_{b2})t}{\sigma\delta_0^2}\right]^{-1/(3+2c_{b2})}, \qquad \frac{\delta}{\delta_0} = \left[1 + \frac{2\nu_0(3+2c_{b2})t}{\sigma\delta_0^2}\right]^{(1+c_{b2})/(3+2c_{b2})}.$$

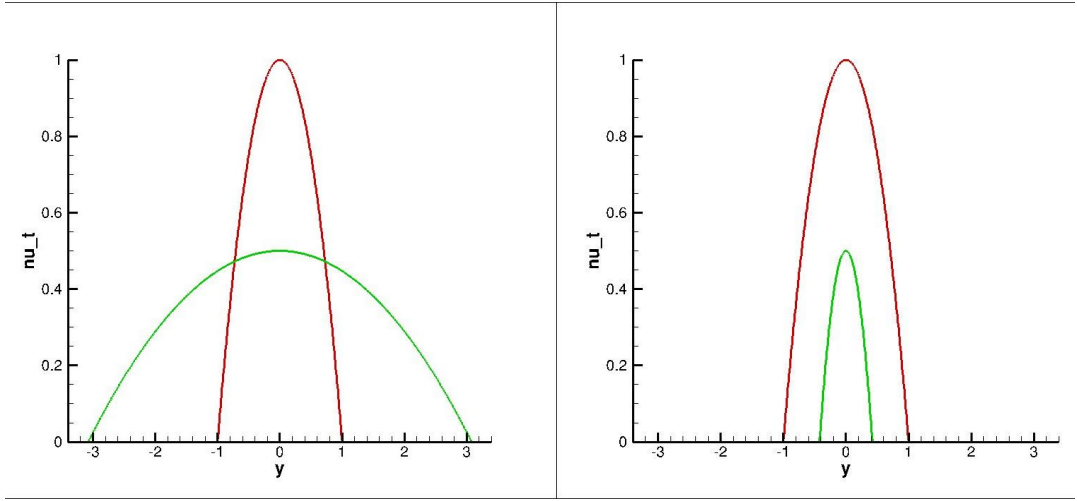

Fig. 3. Parabolic solutions for SA and Baldwin-Barth models. Red, initial state; green, later state.

The importance of the expression $3+2c_{b2}$ (positive for SA, negative for BB) is obvious in these results, but would be very difficult to anticipate from examining the equations, as the SA and BB equations have the same form in their diffusion terms (once BB is re-written with the $c_{b2}$ expression). The SA model obeys the physically expected behavior: the peak eddy viscosity decreases, while the turbulent layer widens. This is realistic, although of course a soft edge would be more realistic; this issue is shared by all common models. In contrast, with the Baldwin-Barth model, the eddy-viscosity distribution shrinks in both directions, and collapses to zero in finite time $t = -\sigma \delta_0^2/(2\nu_0(3 + 2c_{b2}))$. This is not realistic.

We cannot insist enough on the importance of the ETR in any new model that will address external flows, and our purpose here was to show that the mathematics needed is not very difficult. A key helpful fact is that, near the ETR, the diffusion terms dominate.

*Soft Constraints*

We again begin with a list for clarity, but these constraints being somewhat debatable, longer comments will be in order.

1. Do not use wall quantities, such as the friction velocity, at a field point.
2. Do not use boundary-layer quantities such as thicknesses.
3. Do not ignore the Law of the Wall, or the flat-plate boundary-layer problem.
4. Avoid using the molecular viscosity, away from walls. The model must have a well-understood "infinite Reynolds number" behavior.
5. Avoid high derivatives, of the velocity field or the turbulence quantities.
6. Avoid functions, such as min or absolute value, which have discontinuous derivatives. Avoid steep or especially singular behavior of the turbulence variables at the wall.
7. Control the growth of turbulence variables in the irrotational region approaching a stagnation point.
8. Avoid the wall-normal vector.
9. Avoid the von-Karman length scale.
10. Do not use the absolute value of a "legitimate" turbulence input.
11. Provide a Fully Turbulent "safe mode" in which all boundary layers become turbulent, usually by a simple setting of the ambient (inflow) values of the turbulent variables.
12. Provide an "entry point" to build an HRLM around the basic RANS model.
13. Do not propose a model with an unreasonable operation count, such as one with a large Neural Network.

We now explain these constraints. Item 1 is a result of the extreme inconvenience in a solver of needing information from a non-neighbor point. Therefore, although $y^+$ would appear to be a very helpful input, it is avoided and substitutes are used, such as $\chi$ in the SA model. Granted, the wall distance $d$ comes from a non-neighbor point, but it is calculated once and for all.

Item 2 is similar: interpolating the flow field to wall-normal lines is difficult, and the non-neighbor issue is extreme. A recovery for merging shear layers and non-boundary-layer situations is also needed.

Item 3 was discussed at length above.

Item 4 flows from TC, and the dominance in the Reynolds stresses of the large eddies, which do not "feel" the viscosity and drive a one-way cascade of energy and information towards the smaller and ultimately the viscous eddies.

Items 5 and 6: such flaws are an obstacle to high-order accuracy, and they also impede linearization.

The Stagnation-Point Anomaly in Item 7 has been known for many years and occurs because the time for a particle to reach the wall is infinite, and the production is excessive when using the Boussinesq approximation. In addition to stagnation points, it is present at the attachment line of a swept wing. A common solution is to scale production with vorticity instead of strain rate, or to limit it in another way [8].

Some reputable colleagues will disagree with Item 8. There is no question that the wall-normal direction contains crucial physical information, but this vector is discontinuous in space, so that the equations need to ensure that its influence has become negligible before the discontinuity is encountered. A common example is an inside corner, which can be part of a step or a wing-body junction.

Item 9 concerns the quantity which in a simple shear flow reduces to the von-Karman length scale $L_{VK} \equiv |U_y|/|U_{yy}|$; in the log layer, it can be a substitute for $y$. Other length scales derived from the turbulence quantities have been used similarly. The key point is that, if the ETR is at $y = \delta$, in the behavior of classical turbulence models, any length scale falls to 0 and is proportional to $\delta - y$. The vast majority of models use the inverse of the von-Karman scale, and therefore $1/L_{VK} \to \infty$. This will cause problems, and again can be hidden by coarse grids. Besides, $L_{VK}$ is infinite at an inflection point.

This is closely related to a very common issue in "wall-distance free" versions of the SA model (WDF). In these versions, we have in simple terms the following phenomenon, involving production, diffusion and destruction: $D\nu_t/Dt = Prod + Diff - Dest$. In a shear flow driven by $U(y)$ the SA model has $Dest \propto (\nu_t/y)^2$, whereas the WDF model has $Dest \propto (\partial \nu_t/\partial y)^2$. In the log layer, with $\nu_t = u_\tau \kappa y$, both models have $Diff \propto (\partial \nu_t/\partial y)^2$. $Prod$ is positive and is offset by $Diff - Dest$<0 so that $D\nu_t/Dt \approx 0$. Now since the eddy-viscosity distribution across the BL is approximately an inverted parabola, in the outer part of the boundary layer, $\partial \nu_t/\partial y$ is again approximately constant. For normal turbulent entrainment to occur, the model must give $D\nu_t/Dt > 0$. This is ensured in the SA model by the destruction term being much smaller than in the log layer, since $y$ is larger. The WDF model has $D\nu_t/Dt = Diff - Dest$<0 just like before, and therefore the "turbulent front" propagates towards the wall, which is unphysical, as in Fig. 3b. This behavior can be hidden by simulations with too coarse a grid; this motivates the demand for solutions of the flat-plate boundary layer with very fine grids, such as those used at the TMR [8].

Item 10: it is amazing how many models use, for instance, the absolute value of Pope's vortex stretching measure, as if vortex stretching and shortening were equivalent [19]. We have seen a few examples of the same paradox with streamline curvature (as if convex and concave curvature had the same effect), and pressure gradient (favorable and adverse).

Item 11: Fully Turbulent mode is the default for most CFD simulations [18]. Very many engineering boundary layers have only very limited laminar regions, and most often, these are not known, or are far from easy to predict from equations.

Item 12: The essence of an HRLM is to reduce the eddy viscosity relative to the RANS model. This is possible by reducing production or increasing destruction. In two-equation models, the options include at least the production and dissipation of $k$, the same for $\epsilon$ or $\omega$, and the constitutive relation.

Item 13 has been a matter of debate in colleagues' comments. A large NN which depends, for instance, on 100 or 200 exponentials does not appear justifiable; a recent paper has 6 hidden layers with 30 nodes each. On the other hand, the operation count of a seven-equation model must also be heavy, and as mentioned, the slowing down of iterative convergence is a powerful factor.

To summarize, if a model violates a Hard Constraint, it must not be used. If it violates a Soft Constraint, the potential coder/user should hesitate, and ask questions.

## Outlook

It will be excellent if and when ML works in turbulence modeling, in ways small or large. As of 2022, the author has not identified a definite path to that and is more adept at saying what not to do than what to do. He is of course writing within a system of thought that could have gone from old-fashioned to obsolete, and he could be thought to be motivated by a fear of competition. The impact this paper will have is uncertain.

Yet after attending the June 2022 NASA symposium, many observers believe that major obstacles stand in the way of ML for turbulence modeling and have often been simply ignored by the ML contributors [10]. Several of the speakers directly violated one or more of the Hard Constraints, and the same is true for many papers in good journals. This comes in addition to the widespread lack of drive to create a general model. The symposium also failed to give any impression that an ML "community" is taking shape, with a common language, fruitful exchanges, and a down-selection of the better approaches. More will be learned from the final report of the European HiFi-TURB program, in December 2022 [24].

What seems certain is that the idea some of us had of a straightforward combination of large datasets and ML creating useful new models or model versions in a strongly "data-driven" manner will not become reality. Too many of the Constraints presented here (with the ETR a prime example) can be determined only via mathematical analysis of the PDEs by a human. The Mission also requires major non-trivial decisions. Another consideration is the Structural Limitation of classical RANS models the author is attempting to clarify and publish [23]. This could well derail ML exercises in a case as simple as channel flow (assuming the goal is to accurately predict all four Reynolds stresses). A tight team of ML experts and experts with Turbulence Culture will be needed; this is precisely how HiFi-TURB was envisioned [24]. Turbulence research will advance only through rigorous thinking, peer-reviewed publications, and sincere collaboration.

A further remark concerns the future value of RANS modeling, at a time when HRLMs are promoted enthusiastically, and there is much justified pessimism over the ultimate accuracy potential of RANS. The point is that its advantages over HRLM in industrial practice are massive: no residual error from time-averaging; grid convergence; inverse design and optimization; lower computing cost; lower skill level required. Another consideration is that although community efforts and workshops aim at absolute accuracy, this is not indispensable. Wind tunnels are used in industry, accepting many issues CFD is free of, namely: scaling down often to 5% size; confinement; inaccurate aeroelastic deformation; unrealistic size of the slat and flap supports; freestream turbulence. Once a stable "automatic" CFD process with negligible numerical error is established, it will be tested in depth and its inaccuracies be taken into account in Company Culture and Engineering Judgment. The maximum lift coefficient and the buffet boundary will not be exact, but stable corrections will be applied just like they are for wind-tunnel work. If so, the pure accuracy of the turbulence model is of great value, but allowing the fully-converged and repeatable operation of CFD also is.

We end with an informal metaphor. Imagine a ML system has access to a large number of chess games and the knowledge of the reward, namely a checkmate, in a setting of Reinforcement Learning. It has Big Data. However, it is not given all the rules for the motion of the pieces, for instance the knight. When tested after training, the ML product can immediately make a knight perform queen moves, which would be a great advantage but is unacceptable. The chess move rules would give a true Framework. In Turbulence Modeling, we theoreticians are in no position to give the ML systems all the rules. The Hard Constraints listed above are only a start, and besides there is no doubt that some people will ignore some of them. In fact, the expectation that given the right Framework, RANS modeling could truly succeed has nothing to support it. We can only improve the models.

## Acknowledgements

This paper was invited by Prof. R. Dwight and printed in the ERCOFTAC Bulletin 134 of March 2023 with help from T. Buchanan. I enjoyed comments from M. Strelets and his team, F. Menter, C. Rumsey, G. Coleman, T. Knopp, M. Visonneau, A. Smits, and K. Sabnis.

[3] Alauzet, F., Clerici, F., Loseille, A., Tarsia-Morisco, C., and Vanharen, J., "Some progress on CFD high lift prediction using metric-based anisotropic mesh adaptation," AIAA-2022-0388, 2022.

[4] Tracey, B. D., Duraisamy, K., and Alonso, J. J., "A Machine Learning Strategy to Assist Turbulence Model Development," AIAA-2015-1287, 2015.

[5] Ling, J., Kurzawski, A., and Templeton, J., "Reynolds averaged turbulence modelling using deep neural networks with embedded invariance," *J. Fluid Mech*. **807**, 2016, pp. 155-166.

[6] Duraisamy, K. Iaccarino, G., and Xiao, H., "Turbulence Modeling in the Age of Data,. Ann. Rev. Fluid Mechanics **51**, 2018, pp. 357-377.

[7] Wu, J.-L., Xiao, H., and Paterson, E., "Physics-Informed Machine Learning Approach for Augmenting Turbulence Models: A Comprehensive Framework." Phys. Rev. Fluids **3**, 074602 – Published 10 July 2018.

[8] Rumsey, C. L., "The Langley Research Center Turbulence Modeling Resource," https://turbmodels.larc.nasa.gov. Accessed: 2022-09-12.

[9] Rumsey, C. L., "The 4th AIAA CFD High Lift Prediction Workshop (HLPW-4)" https://hiliftpw.larc.nasa.gov, Accessed: 2022-09-12.

[10] NASA Langley Research Center, 2022 Symposium on Turbulence Modeling: Roadblocks, and the Potential for Machine Learning, 27-29 July 2022, Suffolk VA, https://turbmodels.larc.nasa.gov/Turb-prs2022, accessed 2022-08-31.

[11] Rumsey, C.L., Carlson, J.-R., Pulliam, T. H., and Spalart, P. R., "Improvements to the Quadratic Constitutive Relation Based on NASA Juncture Flow Data," AIAA J. **58**, 10, 2020, pp. 4374-4384

[12] Slotnick, J. P., "Integrated CFD Validation Experiments for Prediction of Turbulent Separated Flows for Subsonic Transport Aircraft," NATO Science and Technology Organization, Meeting Proceedings RDP, STO-MP-AVT-307, 2019.

[13] Spalart, P. R., Jou, W. H., Strelets, M., Allmaras, S. R., 1997. "Comments on the feasibility of LES for wings, and on a hybrid RANS/LES approach." In: Proc. 1st AFOSR Int. Conf. on DNS/LES, Greyden Press.

[14] Durbin, P. (ed) Advanced approaches in turbulence, Chapter 4. Elsevier, 2021.
25